\documentclass[aps,twocolumn,pra,aps,showpacs,amsmath,superscriptaddress]{revtex4-1}

\usepackage{epsfig,amssymb,amsmath,amsthm,amsfonts,amsbsy,mathrsfs,amscd}
\usepackage{graphicx}
\usepackage{color}

\usepackage[colorlinks,
            linkcolor= red,
            anchorcolor= blue,
            citecolor= blue
            ]{hyperref}

\usepackage{multirow}
\usepackage{booktabs}
\usepackage{lipsum}

\usepackage{physics}
\usepackage{tikz}
\usepackage{mathdots}
\usepackage{yhmath}
\usepackage{siunitx}
\usepackage{array}
\usepackage{gensymb}
\usepackage{tabularx}

\usepackage{makecell}

\begin{document}
\title{Multi-qubit State Tomography with Few Pauli Measurements}

\author{Xudan Chai}
\affiliation{Beijing Academy of Quantum Information Sciences,
Haidian District, Beijing 100081, P.R. China }

\author{Teng Ma}
\affiliation{Beijing Academy of Quantum Information Sciences,
Haidian District, Beijing 100081, P.R. China }

\author{Qihao Guo}

\author{Zhangqi Yin}
\affiliation{Center for Quantum Technology Research and Key Laboratory of Advanced Optoelectronic Quantum Architecture and Measurements (MOE), School of Physics, Beijing Institute of Technology, Beijing 100081, China}

\author{Hao Wu}

\author{Qing Zhao}

\email{qzhaoyuping@bit.edu.cn}
\affiliation{Center for Quantum Technology Research and Key Laboratory of Advanced Optoelectronic Quantum Architecture and Measurements (MOE), School of Physics, Beijing Institute of Technology, Beijing 100081, China}

\begin{abstract}
In quantum information transformation and quantum computation, the most critical issues are security and accuracy. These features, therefore, stimulate research on quantum state characterization. A characterization tool, Quantum state tomography, reconstructs the density matrix of an unknown quantum state. Theoretically, reconstructing an unknown state using this method  can be arbitrarily accurate. However, this is less practical owing to the huge burden of measurements and data processing for large numbers of qubits. Even comprising an efficient estimator and a precise algorithm, an optimal tomographic framework can also be overburdened owing to the exponential growth of the measurements. Moreover, the consequential postprocessing of huge amounts of data challenges the capacity of computers. Thus, it is crucial to build an efficient framework that requires fewer measurements but yields an expected accuracy. To this end, we built a tomography schema by which only a few Pauli measurements enable   an accurate tomographic reconstruction. Subsequently, this schema was verified as efficient and accurate through numerical simulations on the tomography of multi-qubit quantum states. Furthermore, this schema was proven to be robust through numerical simulations on a noisy superconducting qubit system. Therefore, the tomography schema paves an alternatively effective way to reconstruct the density matrix of a quantum state owing to its efficiency and accuracy, which are essential for quantum state tomography.

\end{abstract}

\maketitle

\section{Introduction}
Recently, explosive advancements in the fundamental research and technological field of quantum physics are attributed to a surge in theoretical studies on entanglement \cite{RevModPhys.81.865, PhysRevLett.125.040502, PhysRevLett.124.062001}, superposition \cite{Bouwmeester2000}, and interference \cite{PhysRevLett.125.020405} as well as technical improvements in precise quantum manipulations \cite{RevModPhys.71.S253} and characterization of quantum circuits \cite{PRXQuantum.2.010201}. These ongoing studies also stimulate the rapid development of quantum information transformation \cite{doi:10.1119/1.1463744, bennett1995quantum}, quantum computing \cite{gill2021quantum,divincenzo1995quantum, ThePhysicsof, Quantuminformation}, quantum cryptography \cite{Pirandola:20, bennett1992experimental}, and quantum simulation \cite{PhysRevLett.125.010501, georgescu2014quantum}. Accuracy and security of quantum states are the most concerning issues in the case of quantum information processing. Therefore, considerable attention is focused on the whole quantum information transformation and quantum computing circuits. Their unit operational blocks are the quantum states on which quantum measurement \cite{PhysRevD.32.3208, SymmetricInformationally} acts and quantum gates where quantum operations can be performed at specific quantum states. Theoretically, a quantitatively certified quantum state could ensure the accuracy and security of quantum information transformation and quantum computing.

Previous studies have discovered a tool to certify the quantum state and reconstruct the density matrix \cite{RevModPhys.29.74, PhysRevA.61.010304} of a quantum state with measurements, namely, quantum state tomography \cite{PhysRevLett.109.120403, lu2015quantum,Measurementofqubits,PhotonicStateTomography,Blume-Kohout_2010,PhysRevA.61.010304,Lukens_2020,4767596}. These studies involved physical processes with an estimation. Physically, extracting information from quantum states results in a large number of measurements. However, mathematically, a valid estimator combined with a recovery algorithm is necessary to post-process the obtained data. This framework suffices to characterize the quantum state in theory precisely. However, tomography reaches its bottleneck owing to the exponentially increasing measurement requirements for qubit multiplications. Accordingly, the huge amount of obtained data might overburden the computer capacity. Moreover, accomplishing a full tomographic scan for qubits larger than 10 will be technologically difficult. Therefore, a more efficient method is urgently needed, which requires fewer measurements but still yields high accuracy. Linear inversion \cite{doi:10.1190/1.1444033}, maximum likelihood estimation (MLE)\cite{https://doi.org/10.1111/j.2517-6161.1961.tb00430.x}, Bayesian mean estimation (BME)\cite{PhysRevA.85.052120}, compressed sensing (CS) \cite{Flammia_2012, gross2010quantum}, Fisher information \cite{ma2011quantum} and the self-guided method \cite{PhysRevLett.113.190404, chapman2016experimental} are common estimators. For example, the CS can reduce the measurement requirements owing to its compressed sampling. Recovery algorithms also include CVX \cite{ma2011quantum}, a two-step descent method, and Nesterov \cite{PhysRevA.99.042321}, a one-step descent method. Additionally, the Powel \cite{powell1978fast}, APG \cite{li2015accelerated}, CG \cite{hanke1997regularizing}, and APG-CG  \cite{PhysRevA.95.062336}  are used to promote accuracy and speed.

\begin{table*}
\small
\centering
\setlength{\tabcolsep}{0.2mm}{
\begin{tabular}{c|c|c|c|c} 
\toprule[1pt]
qubit & Real part (P) & Imaginary part (P) & Bases (P) & Bases (F) \\
\hline
1 & X,Z & Y&3 & 3 \\
\hline
2&XX,ZZ,YY&YZ,ZY&5 & 9\\
\hline
3&XXX,ZZZ,YYZ&YZZ,ZYZ,ZZY&6 & 27\\
\hline
4&XXXX,ZZZZ,YYYY&YZZZ,ZYZZ,...,ZYYY,...,YYYZ&11 & 81\\
\hline
5&XXXXX,ZZZZZ,YYYYZ&YZZZZ,ZYZZZ,...YYYZZ,YYZYZ,...&28 & 243\\
\hline
6 &XXXXXX,ZZZZZZ,YYYYYY&YZZZZZ,ZZZZZY,XXXXXY,YXXXXX,...&135 & 729\\
\hline
$\vdots$ & $\vdots$ & $\vdots$ & $\vdots$ & $\vdots$ \\
\hline
$2n-1$ & $X^{\otimes 2n},Z^{\otimes 2n},Y^{\otimes 2n-1}  Z$ & $Y Z^{\otimes 2n-2}, \cdots , Y  X^{\otimes 2n-2},\cdots$ & \makecell[c]{$3  + A_{2n-1}^{1} + A_{2n-1}^{3} $ \\  $+ \cdots + A_{2n-1}^{3} + A_{2n-1}^{1} $}  & $3^{2n-1}$ \\
\hline
$2n$ & $X^{\otimes 2n},Z^{\otimes 2n},Y^{\otimes 2n}$ & $Y  Z^{\otimes 2n-2}, \cdots, Y  X^{\otimes 2n-2},...$ & \makecell[c]{ $3 + A_{2n}^{1} + A_{2n}^{3}$  \\  $+ \cdots + A_{2n}^{3} + A_{2n}^{1}$ } & $3^{2n}$ \\
\bottomrule[1pt]
\end{tabular}}
\caption{Measurements for reconstructing the density matrices of quantum states of qubits system. The basis measurement law differs between the quantum states with odd ($2n-1$) and even ($2n$) numbers.}  MSE $\sim10^{-2}$ is the stopping condition. The rightmost column is the number of bases for full tomography $\mathrm{[Base \ (F)]}$, compared with that for our method $\mathrm{[Base \ (P)]}$. 

\label{table1}
\end{table*}
The aforementioned studies \cite{, PhysRevA.106.012409, PhysRevLett.130.090801} attempted to construct an efficient and accurate tomographic framework. These studies are becoming more important in the case of a large number of qubits which is required in practical quantum computation and quantum information processing. Here, we construct a tomographic schema that requires fewer Pauli measurements \cite{doi:10.1063/1.528006} but still yields high accuracy. Theoretically, this work is rooted in the previous work \cite{PhysRevA.99.042321}, in terms of picking Phaselift as the estimator and choosing Nesterov as the recovery algorithm. However, the goal of this work differs from that of \cite{PhysRevA.99.042321}.  The schema here has been proven to be valid and accurate through numerical simulations in multi-qubit state tomography. In particular, a state without a phase factor \cite{PhysRevLett.58.1593} can be precisely reconstructed by measuring only two Pauli bases.

\section{Method}
\subsection{Theoretical analysis}
In general, a density matrix for an $n$-qubit system can be represented by the Stokes parameters  Generally, a density matrix for a n-qubit system can be represented by the Stokes parameters:

\begin{equation}\label{stoke}
\rho_n=\frac{1}{2^n} \left( \sum_{ \mathbf{u} = 0  }^3 c_\mathbf{u} \ \sigma_{  \mathbf{u} }  \right),
\end{equation}
where and $c_\mathbf{u} \in \mathbb{R}$, with $\textbf{u}=i,j,\cdots,k$, denotes  the Stokes parameter, and $\sigma_\mathbf{u}:=\sigma_i\otimes \sigma_j \otimes \cdots \otimes  \sigma_k$ in which
 $\sigma_{0,1,2,3}$ corresponds to  the identity matrix and the Pauli matrices: 
\begin{equation*}  
\begin{aligned} 
& \sigma_0 := I := \left(              
\begin{array}{cc}   
1 & 0\\  
0 & 1\\  
\end{array}
\right) ,   \,
 && \sigma_1 := \sigma_x \equiv\left(             
\begin{array}{cc}   
0 & 1\\  
1 & 0\\  
\end{array}
\right)   , \\
 \, 
& \sigma_2 := \sigma_y  \equiv\left(                
\begin{array}{cc}   
0& -i\\  
i& 0\\  
\end{array}
\right), 
&& \sigma_3 := \sigma_z  \equiv\left(                
\begin{array}{cc}   
1&0\\  
0&-1\\  
\end{array}
\right),   
\end{aligned}       
\end{equation*}
respectively. A Stokes parameter $c_\mathbf{u}$ can be obtained using Pauli measurement setting $\textbf{u}$, in which we meant measuring the observable $\sigma_\mathbf{u}$ was measured under the tomographic state. 
\begin{equation}
	 c_\mathbf{u} = \textrm{tr} ( \sigma_{  \textbf{u} } \rho_n  ) =  p_\mathbf{u}^+ - p_\mathbf{u}^- \, ,
\end{equation}
where the probabilities 

\begin{equation}
p_\mathbf{u}^\pm  = \sum_{ \mathcal{M}_{\mathbf{u},\mathbf{b} } \in \mathcal{S}_{\mathbf{u}}^\pm} \tr (\mathcal{M}_\mathbf{u,b} \,  \rho_n),
\end{equation}
 with $\mathcal{S}_\mathbf{u}^{\pm}$ ± denotes the eigen subspace of $\sigma_{\mathbf{u}}$ with the eigenvalues $\pm 1$, and $\mathcal{M}_{\mathbf{u,b}}$ denotes a projector for an eigen product basis of $\sigma_\mathbf{u}$.      
Here note the subscripts $0,1,2,3$ in Eq. (\ref{stoke}) correspond to $I, X, Y, Z$ measurements, respectively, moreover, we say $X, Y, Z$, instead of $1,2$ and $3$,  measurement setting  hereinafter.

All the diagonal elements of the density matrix can be determined via the $Z$ direction measurement, which used $\sigma_z$ combined with $\sigma_0$. All the measurement settings  $U \cdots V$ with $U, \cdots, V \in \{X, I\}$ can share a common set of  measurement product basis, from which the Stokes parameters $c_{U\cdots V}$ can be obtained from a common set of probabilities. Moreover, for the real part of the density matrix, because all the Stokes parameters with the odd number of $Y$ are zero, only the measurements with an even number of $Y$ are needed.  Similarly, for the imaginary part of the density matrix, only the measurements with the odd number of $Y$ are needed.

Based on the aforementioned analysis, we propose the measurement shown in Table \ref{table1} for state tomography for a class of entangled pure states including GHZ, W, and cluster state. For the real part (without a phase factor), a $Z$ direction measurement is necessary to determine the diagonal elements, and an even number of $Y$ together with $X(Z)$ direction measurements are used to reconstruct the coherent (off-diagonal) elements. For the imaginary part, odd numbers (smaller than the number of qubits) of $Y$ together with $Z$ or $X$ are necessary to reconstruct the off-diagonal elements. The specific forms of the measurements shown in Table \ref{table1} are verified through our numerical simulation  in later sections.   

We also propose more general law in Table table1 to reconstruct the pure or nearly pure state, by considering an $n$-qubit (with the number of qubits $n>6$) state, where the situations are different for $n$ is odd or even. The case is simpler when n is even. Here, reconstructing the real parts requires the measurements $X^{\otimes n}$, $Y^{\otimes n}$, and $Z^{\otimes n}$, while $Y^{\otimes n}$ measurements become unnecessary for states without phase factors, i.e., all the elements of the density matrix are real. When $n$ is odd, the measurement $Y^{\otimes n-1} \otimes Z$ or $Y^{\otimes n-1} \otimes X$ is required.

As for the limitation of our method, it should be pointed out that the measurements presented in Table \ref{table1} are not sufficient to uniquely determine an arbitrary pure state among all states. Since it has been proven in \cite{ma2016PRAa} that 11 and 31 Pauli measurements are needed  to uniquely determine an arbitrary 2- and 3-qubit pure state among all states, respectively. 
Moreover,  note that the efficiency of the Nesterov recovering algorithm in \textbf{B} of this Section requires the state should be of low rank. 
Whilst, we find that, through numerical simulation in Section \ref{Section--Simulation}, the Pauli measurements in Table \ref{table1} can determine the specific states including W, GHZ, and cluster state, with high precision and efficiency. 

\subsection{Nesterov Recovering algorithm}
After framing a schema of measurements using which the density matrices can be reconstructed, an efficient estimator combined with a precise recovering algorithm is mathematically required to solve this problem numerically. Herein, we introduce the PhaseLift method \cite{lu2015quantum} as the estimator and the Nesterov algorithm\cite{NESTA} as the recovery core, the validity of which has been demonstrated previously \cite{PhysRevA.99.042321}. Integrating this estimator and recovery core allows us to efficiently recover the elements of the density matrix. Based on the method presented in \cite{PhysRevA.99.042321}, the reconstruction problem can be formulated into a convex optimization problem:

\begin{equation}
\begin{aligned}\label{APGL}
\text{Minimize} \Bigg\{ \sum\limits_{\mathbf{u}, \mathbf{b}} \frac{1}{2} \big[ \textrm{Tr} \left(\mathcal{M}_{\mathbf{u},\mathbf{b}} \,  \rho \right) - f_{\mathbf{u},\mathbf{b}} \big]^2 + c_{1}&||\rho||^{2}_{F} \\ +c_{2}\text{Tr}(\rho) \Bigg\},\quad\text{subject to} \quad \rho \geq 0,
\end{aligned}
\end{equation}
where the relative frequency  $f_{\mathbf{u},\mathbf{b}}$ denotes the actual probability of the measurement result of $\mathcal{M}_{\mathbf{u},\mathbf{b}}$, and $c_{1,2}$ represent optimization parameters in the APGL algorithm \cite{toh2010accelerated}. In particular,   $f_{\mathbf{u},\mathbf{b}}$ is approximately equal to the expected value of ${\mathcal{M}}_{\mathbf{u},\mathbf{b}}$ when the number of measurements is sufficiently large, $c_1$ is related to the upper bound of the Lipschitz constant  and $c_2$ is related to the constrain of the semi-definite property of the density matrix.   

The compressed sensing\cite{gross2010quantum} method has been widely used in quantum state tomography, which only requires spending much fewer Pauli measurement settings than that of the full tomographic method. However, the compressed sensing\cite{gross2010quantum} method does not specify which Pauli measurement set can be used.

The previous theoretical research \cite{PhysRevA.99.042321} on comparing the performances between other methods including CVX, APG, and CG methods has verified the high accuracy and high speed for large numbers of qubits state tomography, especially for specific entangled states, such as W state, cluster state, and GHZ state. 
  
\begin{figure}
\includegraphics[width=9cm]{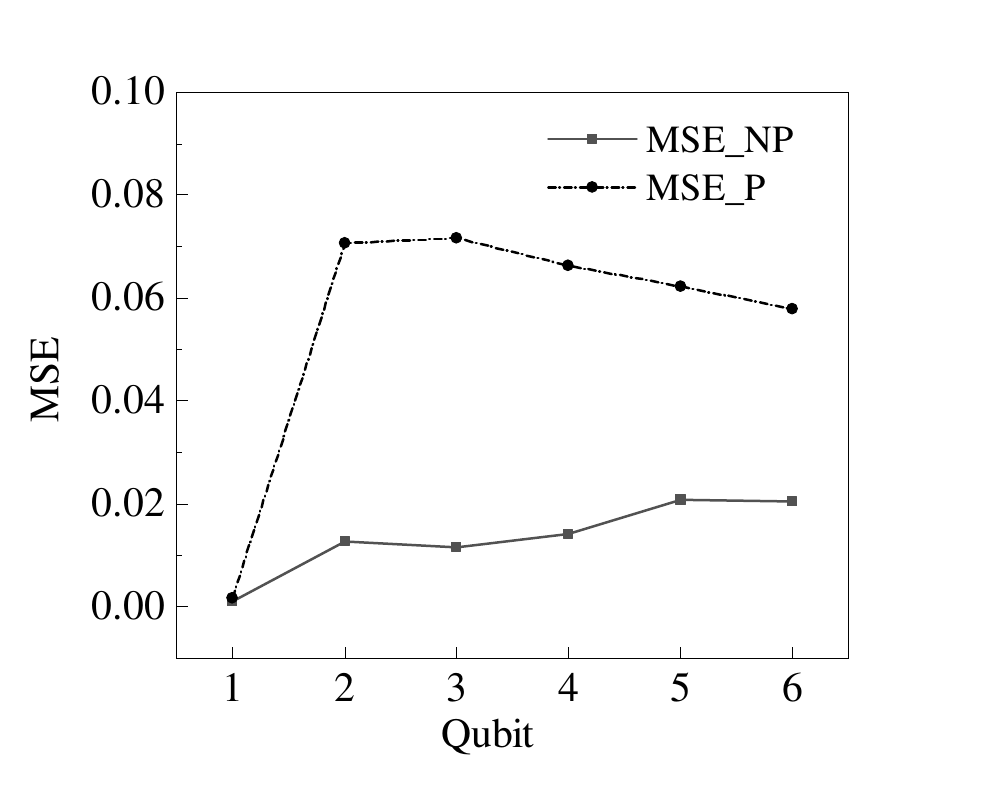}
\caption{\label{Fig2} Schematic for the reconstruction of 1- to 6-qubit pure states using our efficient tomography schema. The dashed line (\textrm{MSE\_P})and solid line (\textrm{MSE\_NP}) represent the states with and without phase factors, respectively.}
\label{pure state}
\end{figure}

\begin{figure*}[t]
\centering
\includegraphics[width=14cm]{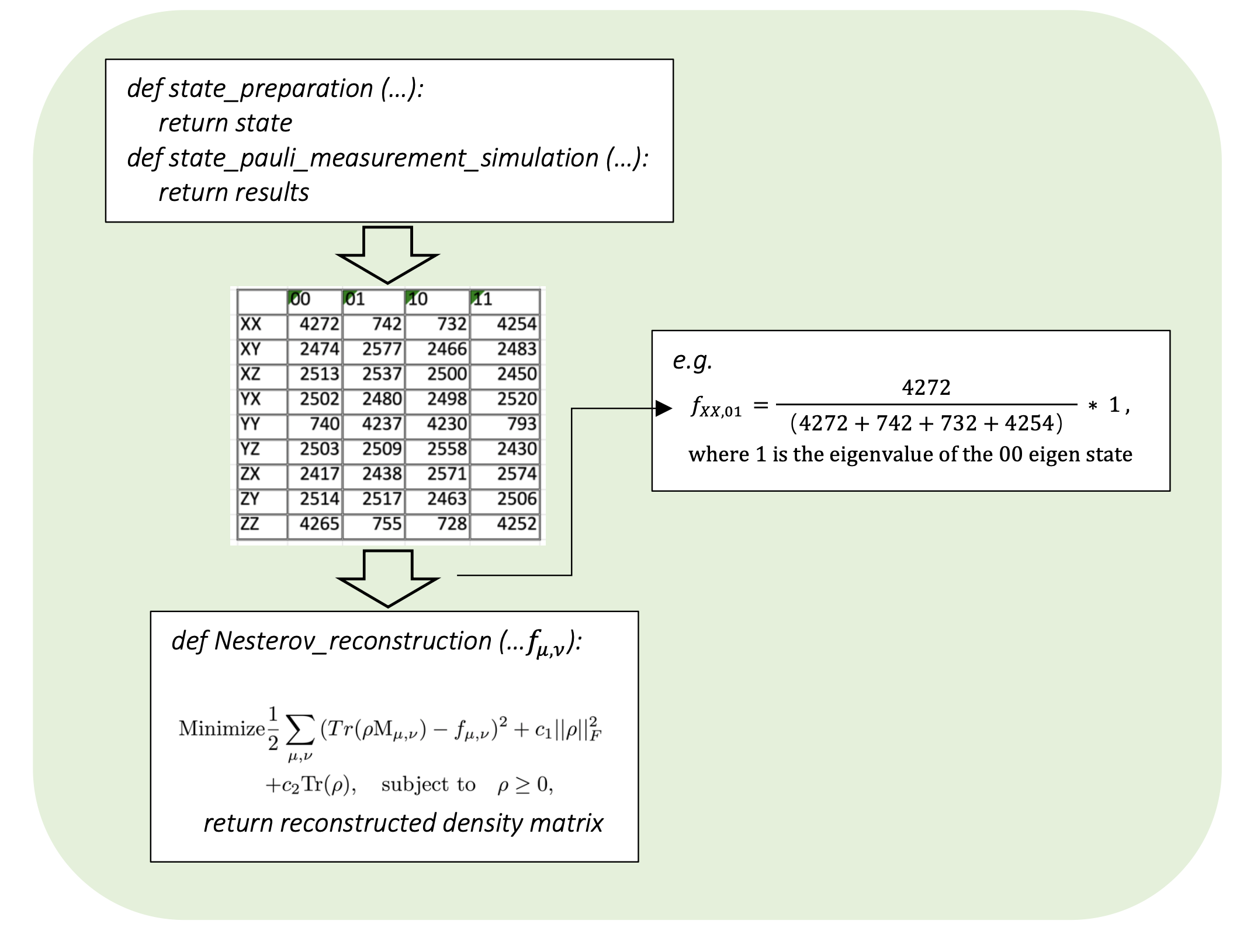}
\caption{\label{Fig3} Schematic diagram for the procedure of the tomographic schema.}
\label{nflow}
\end{figure*}

Recently, IBM has become a very important quantum computation platform in the cloud. We use this platform to simulate and execute real tomography experiments. The tomography methods in $\mathtt{qiskit. ignis.}$ verification of IBM can be categorized into two parts, one part is designed for reconstructing the density matrix of the quantum state, and the other part is for characterizing the performance of a quantum circuit by estimating the average gate fidelity or average error rate. Here we focus on the state tomography in qiskit and we find it includes MLE and CVX method. The density matrix reconstruction can be treated to solve the linear problem, we can turn it into an optimization problem by attempting to minimize the problem while subjecting it to additional constraints to ensure it is indeed a density matrix. This is done by $\mathtt{state_{cvx \ fit}}$. Another approach is to solve this optimization problem with no further constraints. The result might not be a density operator, i.e. positive semi-definite with trace 1; in this case, the algorithm first rescales in order to obtain a density operator. This is done using $\mathtt{state_{mle \ fit}}$.

Therefore, on this platform, we numerically simulated multi-qubit graph state tomography on a noisy superconducting qubit platform in the IBM cloud \cite{PhysRevLett.122.080504} to explore the robustness of this schema. Investigations showed that two Pauli measurements enabled us to reconstruct multi-qubit states \cite{PhysRevA.69.062311}.

\subsection{The procedure of the tomographic schema}
In order to make it clear, we will explain this method in detail. The procedure of the simulation method  (as shown in Fig. \ref{nflow})  is composed of three steps:

Part 1: Get data through numerical simulation.

Part 2: Process the obtained data.
Here we feed the data into the estimator, i.e., Eq. \eqref{APGL}. And, if we want to feed the relative frequency into the estimator, we need to multiply by the corresponding eigenvalues of the state. 

Part 3: Execute the Nesterov algorithm, then we can obtain the estimated density matrix.

Due to the similarity between this work and the previous one\cite{PhysRevA.69.062311}, it is necessary to demonstrate the difference between them.
First, the numerical simulations in  \cite{PhysRevA.99.042321} just prove that  it is possible to do the multi-qubit state tomography accurately  and efficiently in theory, and the Gaussian white noise in those simulations is less practical. Here,  we prove the practical possibility of our method by feeding the real experimental data, which is obtained from the IBM cloud, into the tomographic schema.

Second, we have verified the accuracy and efficiency of the method in \cite{PhysRevA.99.042321}, which shows that only partially randomly selected Pauli bases produce a highly precise multi-qubit tomography.
Nonetheless, for a specific quantum state, the bases are not specified. Hence in this work, we naturally consider that the number of measurement basis might be cut down, especially for some special quantum states, as recent studies have suggested both in theory and experiment \cite{overtomo}. 

In summary, while the previous works have paved the theoretical path, this work uses this method in practice  and finds that fewer measurements can produce a precise  multi-qubit tomography.
\section{Simulations on tomography of multi-qubit quantum states}\label{Section--Simulation}

We have theoretically constructed an efficient schema to reconstruct the density matrix of an unknown quantum state by only measuring only a few Pauli bases. Its validity and efficiency turned out to be true through the following simulations on tomography of multi-qubit pure states. 

Fig. \ref{pure state} shows the simulation of the multi-qubit GHZ state reconstructions with an efficient tomography schema. Hereinafter, we used the mean square error (MSE), $\text{Tr}\left[(\rho_{\textrm{E}}-\rho_{\textrm{T}})'(\rho_{\textrm{E}}-\rho_{\textrm{T}}) \right] $, and fidelity, $\text{Tr}(\rho_{\textrm{E}} \rho_{\textrm{T}})$, to evaluate the accuracy, where $\rho_{\textrm{T}}$ and $\rho_{\textrm{E}}$ denote the target state and the estimated state, respectively. This schema shows higher accuracy and efficiency for the states without phase factors than the states with phase factors compared with the states with phase factors. Therefore, this method is best suited for the tomography of stats that are expected to have no imaginary parts.

\begin{figure*}
\centering
\includegraphics[width=19cm]{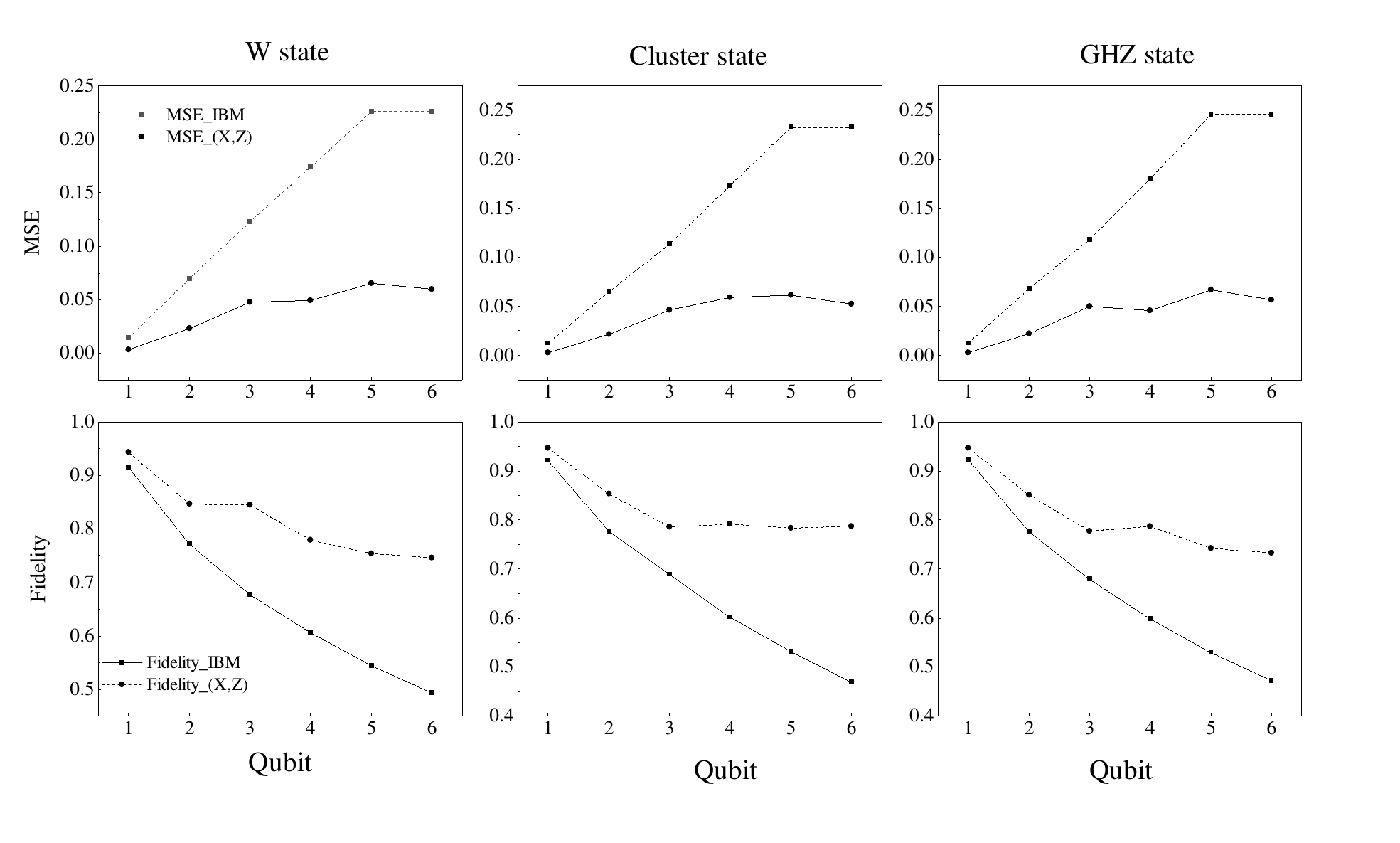}
\caption{\label{Fig3} Tomography simulations for the W, cluster, and GHZ states. The solid lines ($\mathrm{MSE/Fidelity\_{(X,Z)}}$) represent our efficient tomography schema, which only measures two directions, $X$ and $Z$. The dashed lines (MSE/Fidelity IBM) represent the results of the IBM cloud platform. The accuracy and fidelity of the reconstructions are estimated using MSE (shown in the upper and lower panels, respectively).}
\label{graph state}
\end{figure*}

\begin{figure*}
\centering
\includegraphics[width=19cm]{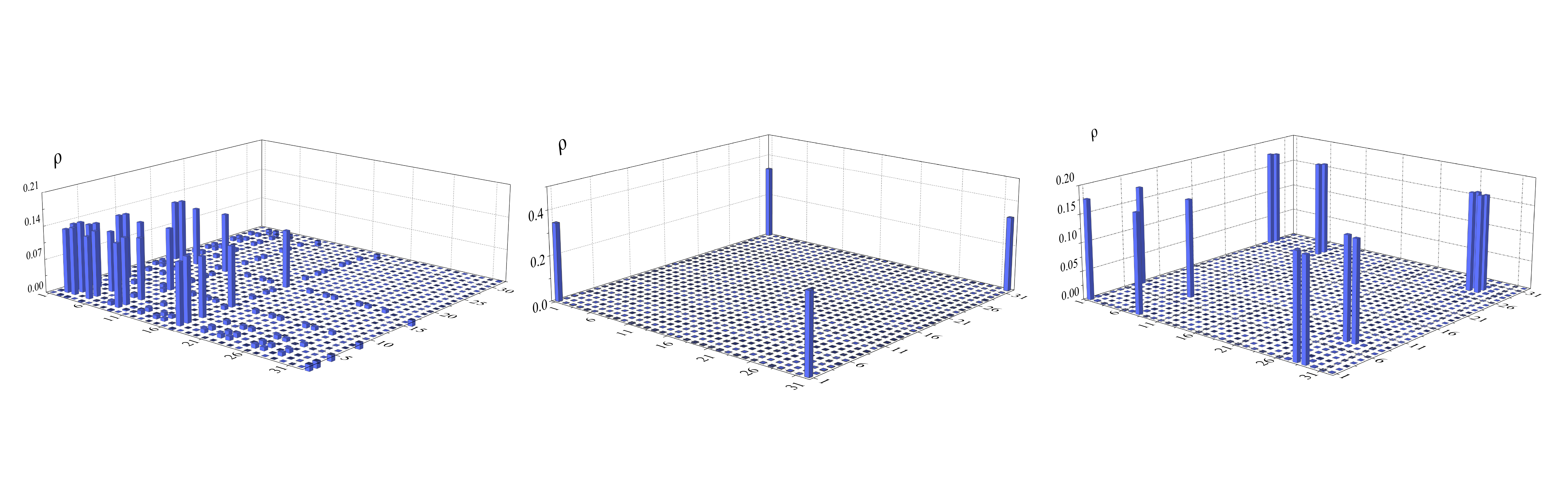}
\caption{\label{Fig4} Populations of density matrices of 5-qubit W, cluster, and GHZ states respectively using our efficient tomography schema.}
\label{fig-density matrix}
\end{figure*}
According to the above simulations, it is necessary to 
test this efficient tomography schema for some specific entangled states. Therefore, we use the IBM quantum cloud \cite{PhysRevLett.122.080504} to simulate the tomography process of quantum states in a superconducting qubit system \cite{barends2016digitized}. Three typical graph states are selected, namely, the W, cluster, and GHZ states, owing to their intensive applications in quantum information and quantum computation. The n-qubit W state and n-qubit GHZ state are expressed as follows:
\begin{equation}
\begin{aligned}
&| \mathrm{W} \rangle_n = \frac{1}{\sqrt{n}} \big(|0_1\cdots0_{n-1}1_n\rangle + |0_1 \cdots 0_{n-2}1_{n-1} 0_n \rangle \\
& \hspace{0.11\textwidth} + \cdots + |1_1 0_2 \cdots 0_n \rangle   \big) , \\
&|\mathrm{GHZ} \rangle_n = \frac{1}{\sqrt{2}} \left( |0\rangle^{\otimes n}+ |1\rangle^{\otimes n} \right) .
\end{aligned}
\end{equation}
We consider the linear type cluster state, which has the exact form,
\begin{equation}\label{cluster_line}
\left|\mathrm{C}\right\rangle_n=\frac{1}{2^{n / 2}} \bigotimes_{i=1}^n \left[|0\rangle_i +|1 \rangle_i  \otimes  \sigma_z^{(i+1)} \right] ,
\end{equation}
with the convention $\sigma_z^{(n+1)}\equiv 1$. 
From Eq. (\ref{cluster_line}), one can have the exact forms of the 2-, 3-, and 4-qubit cluster states: 
$|\mathrm{C}\rangle_2 = (|0+ \rangle + |1- \rangle )/\sqrt{2}$, $|\mathrm{C}\rangle_3 =  (|+0+ \rangle + |-1- \rangle )/\sqrt{2}$, and $|\mathrm{C}\rangle_4 = (|+0+0 \rangle + |+0-1 \rangle + |-1-0 \rangle  + |-1+1 \rangle)/\sqrt{2}$. Notably, the 2-  and 3-qubit cluster states are local unitary equivalence to the GHZ state, whereas any $n>3$ multiqubit  cluster state is not \cite{briegel2001PRL}. In a quantum circuit, the cluster state is generated in accordance with its corresponding graph, in which vertices represent qubits with the initial states $|\pm \rangle \equiv  (|0\rangle \pm |1\rangle)/\sqrt{2}$, and the edges represent the controlled-phase gates acting on the qubits afterward.  


In simulating the real qubits system in practice, we introduce noise models in a quantum circuit. The errors on each quantum gate and the process of reading out are also considered. In particular, we apply a depolarizing channel model to simulate the errors of quantum gates (where the single- and two-qubit gates error rates are set to be $a = 0.002$ and $b = 0.005$, respectively). Moreover, statistical noises are very important in quantum computation, and in our simulation, this kind of noise has been modeled with a bit flip error with a specific probability when the qubits are measured.

Fig. \ref{graph state} shows 1- to 6-qubit  W, cluster, and GHZ state tomography evaluated  by EMS and  fidelity. 
Herein, we reconstructed density matrices using only two measurement settings, $X^{\otimes n}$ and $Z^{\otimes n}$ with $n$ being the number of qubits. Consequently, our tomography schema, with a smaller MSE and higher fidelity, outperformed the IBM cloud with regard to accuracy. Therefore, this schema might be applied to state certifications for large numbers of qubits. Moreover, the schema showed higher efficiency than the IBM cloud platform. 
The reconstructed density matrices are shown in Fig. \ref{fig-density matrix}. 


\section{Discussion on the robustness of the efficient tomography schema}

\begin{figure}
\includegraphics[width=9cm]{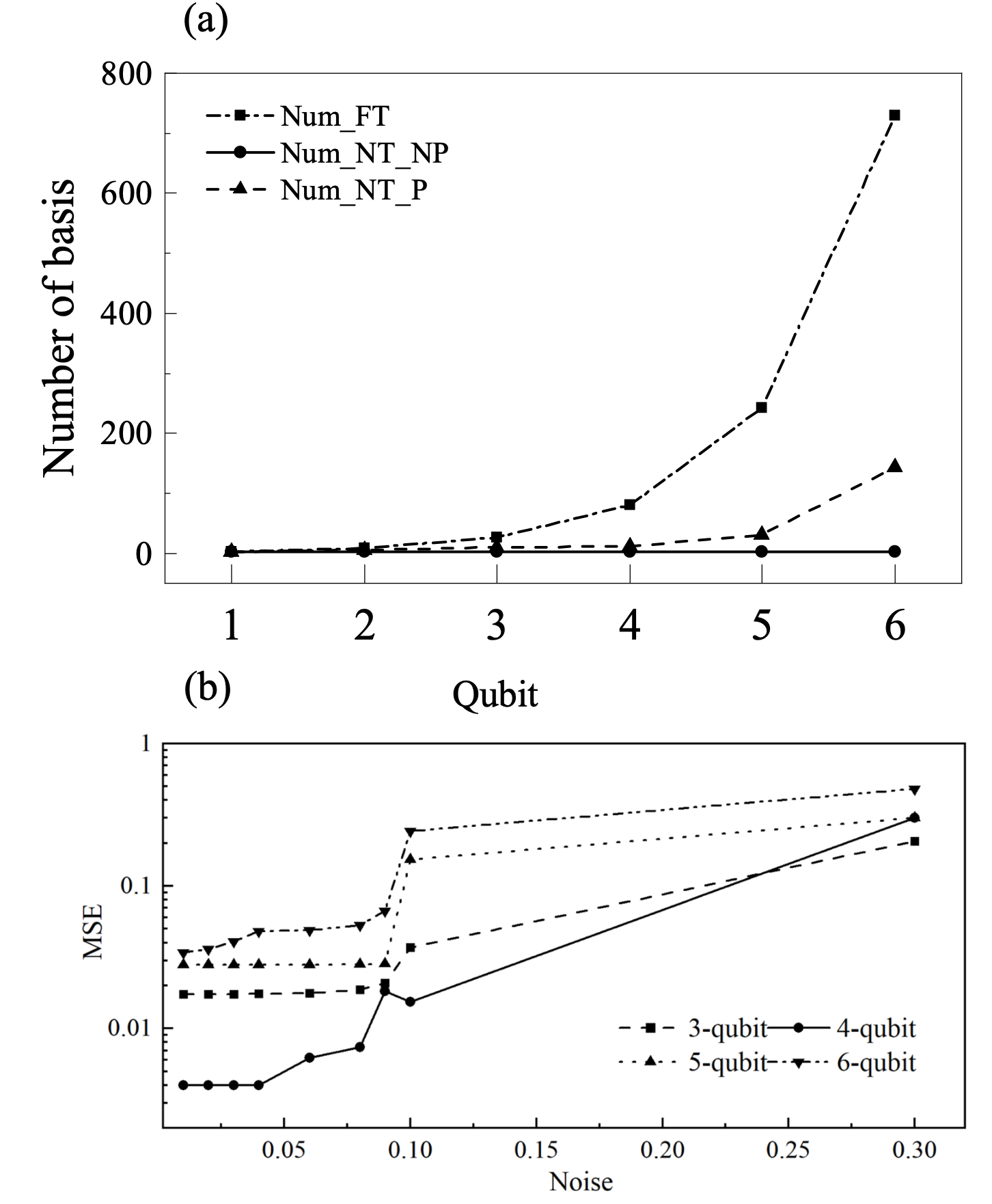}
\caption{\label{Fig4}  (a). Comparison between the required measurements of the full tomography and efficient tomography schema. The dashed-dotted line ($\textrm{Num\_{FT}}$), solid line ($\textrm{Num\_{NT}\_NP}$), and dashed line ($\textrm{Num\_{NT}\_P}$) denote the number of measurements with full tomography, efficient tomography schema without a phase, and efficient tomography with a phase, respectively. (b). Robustness of our efficient tomography schema. The horizontal axes represent the noise level of the two-qubit gate.}
\label{robustness}
\end{figure}

A comparison between the full tomography and efficient tomography schema is shown in Fig.\ref{robustness}(a). We can observe that the number of measurements for the efficient tomography schema increases polynomially, whereas it increases exponentially for the full tomography schema.
In particular, in the absence of a phase factor, two measurements can be used for the reconstruction. Evidently, this method outperforms the full tomography schema in terms of efficiency.
A reliable quantum computing and information process requires a precisely controlled qubit system. The qubit system that interacts with its environment at all times becomes easily incoherent, and the noise from the environment technologically challenges the multiplication of the quantum system. Additionally, noises are generated during state preparation and measurement. Therefore, investigating the robustness of the tomography schema under noise for the multiqubit system is essential.
Fig.\ref{robustness}(b) shows the numerical simulation of multiqubit state tomography for a noisy superconducting qubit system. We investigate the robustness of our tomography schema by varying the two-qubit error rate b from 0 to 0.2 and fixing the single-qubit error rate at a = 0.002 as well as the state-of-the-art noise level of the superconducting qubit system. The simulation shows that the noise has a marginal effect on the state tomography as long as the error rate is small enough, where b is smaller than $~$0.08. This type of robustness remains when the number of qubits multiplies. Therefore, this schema has promising robustness, even if the scale of qubits grows.

It is worthy
to introduce new researches that are demonstrated both
in theory and experiment\cite{overtomo}.
The tomographic schema demonstrated by them enables us to accurately  determine an unknown quantum state with at most  $e^{O(k)} \log^{2}(n)$ (where $n$ is the number of qubits and $k$ is...), which is realized by concrete measurement protocols. The protocol originates from a theory in the quantum computation field which is called perfect hash functions. Using this theory we can partition the qubits into disjoint subsets, in which we perform all parallel measurements over single-qubit measurement bases ($X$, $Y$, or $Z$), such that all qubits in the same subset have identical measurement settings. This procedure allows us to determine all the reduced density matrices using  only $NM3^{k}$ measurements  altogether,  where $N$, $M$, and $k$ are the number of the partitions, the measurements in an individual base, and the disjoint subsets, respectively. With this theoretic base, the experiment later realizes this theoretic schema in a phonic quantum state reconstruction, using MLE and BME.

In summary, the efficiency and accuracy of the tomography methods in the above studies are mainly guaranteed by efficient measurement protocol instead of an estimator or recovery algorithm. Hence the tomographic schema we put forward here utilizes a different estimator, called Phaselift, and implements a Nesterov algorithm,  which realizes the accuracy and efficiency of tomography. 
\section{Conclusion}
We have theoretically framed a tomography schema using which multiqubit states can be reconstructed efficiently and accurately. Physically, the efficiency was ensured using a fewer number of Pauli measurements compared with those required for full tomography. Mathematically, the accuracy of this tomography schema was achieved using a combination of the PhaseLift method (as the estimator) and the Nesterov method (as the recovery algorithm). Our tomography schema outperformed the full tomography in terms of efficiency and accuracy, particularly in the case of a larger number of qubits, as justified by numerical simulations on the tomography of three entangled states, i.e., W, cluster, and GHZ states. Considering a quantum state certification that requires fewer measurements but yields a precise estimation, this efficient tomography schema might be useful in practical quantum computing and information transformation.

\section*{ACKNOWLEDGEMENT}
We would like to thank Bo Gao for his careful revision of the manuscript. This work is supported by the NSFC Grant Nos. 11905100  and 11675014. Additional support is provided by the Ministry of Science and Technology of China (2013YQ030595-3).


%

\end{document}